\documentclass[twocolumn,aps,pra,notitlepage,bibnotes10pt,superscriptaddress,nofootinbib]{revtex4-1}
\usepackage{amsmath}
\usepackage{graphicx}
\usepackage[normalem]{ulem}
\usepackage{dcolumn}
\usepackage{epsfig}
\usepackage{bm}
\usepackage{array}
\usepackage[english]{babel}
\usepackage{hyperref}
\usepackage{algorithm,algorithmic}
\usepackage{float}

\hypersetup{
colorlinks=true,
citecolor=blue,
linkcolor=red,
urlcolor=blue
 ,
 pdfmenubar=true
}

\usepackage{hyperref}
\usepackage{graphicx}
\usepackage{amsmath}
\usepackage{amsfonts}
\usepackage{amssymb}
\usepackage{xcolor}
\usepackage{bbm}
\usepackage{calrsfs}
\usepackage{dutchcal}
\usepackage{amsthm}
\usepackage{tikz}
\usetikzlibrary{calc,arrows,chains,matrix,positioning,scopes}

\AtBeginDocument{%
    \newwrite\bibnotes
    \def\bibnotesext{Notes.bib}
    \immediate\openout\bibnotes=\jobname\bibnotesext
    \immediate\write\bibnotes{@CONTROL{REVTEX41Control}}
    \immediate\write\bibnotes{@CONTROL{%
    apsrev41Control,author="08",editor="1",pages="1",title="0",year="1"}}
     \if@filesw
     \immediate\write\@auxout{\string\citation{apsrev41Control}}%
    \fi
}%

\newcommand{\bra}[1]{\ensuremath{\langle#1|}}
\newcommand{\ket}[1]{\ensuremath{|#1\rangle}}

\newcommand{\ketbra}[1]{\ensuremath{| #1 \rangle \langle #1 |}}

\newcommand{\be}{\begin{equation}}
\newcommand{\ee}{\end{equation}}

\newcommand{\beq}{\begin{eqnarray}}
\newcommand{\eeq}{\end{eqnarray}}

\begin{document}

\title{Creating nonlocality using geometric phases between partially distinguishable photons}
\author{Valentin~Gebhart}
\affiliation{QSTAR, INO-CNR and LENS, Largo Enrico Fermi 2, 50125 Firenze, Italy}

\begin{abstract}
The geometric (Berry-Pancharatnam) phase originates from the intrinsic geometry of the space of quantum states and can be observed in different situations, such as a cyclic evolution of a quantum system. 
Here, we utilize the geometric phase to obtain a surprising insight: It is possible to create nonlocal correlations in a fixed interferometer with independent photon inputs by varying the photons' internal states.
In particular, we consider a cyclic interferometer that is fixed, i.e., that has no variable internal phase shifts or subsequent measurement settings.
Instead, the measurement choices of the different parties correspond to the internal states of the input photons which influence the observed correlations via a collective $N$-photon geometric phase, constituting a different approach for the generation of nonlocality with respect to the usual paradigm. 
We observe a trade-off between the geometric phases and the visibility of the many-photon interference, impeding the generation of nonlocality. 
However, by making use of the dynamical quantum Zeno effect, we show that nonlocality can be created in the fixed cyclic interferometer using $12$ (or more) independent photons. 
\end{abstract}

\maketitle

\section{Introduction}
The geometry of quantum state space can reveal itself in observable quantities such as geometric phases~\cite{berry1984,pancharatnam1956,wilczek1989,chruscinski2004geometric}. 
Most prominently, a geometric (Berry) phase was identified for quantum systems that undergo a closed adiabatic evolution~\cite{berry1984}. 
This phase is termed geometric because it only depends on the trajectory of the quantum state and not on the speed (or the energy) of the state's evolution: 
The phase is determined by the solid angle in projective Hilbert space that is enclosed by the state's trajectory. 
Over the years, the geometric phase and related notions have shown to be connected to several quantum effects~\cite{wilczek1989,xiao2010,chruscinski2004geometric} such as, e.g., topological phases of quantum matter~\cite{thouless1982,bernevig2013,asboth2016} and the fractional statistics of anyons~\cite{wilczek1990,law2006}, and they have been employed for different quantum information processing tasks~\cite{ekert2000,jones2000,falci2000,zhu2005,nayak2008}. 
Geometric phases not only occur in a closed adiabatic quantum evolution, but also in a general unitary evolution~\cite{aharonov1987} that need not to be closed~\cite{samuel1988}. 
Furthermore, the geometric phase can also be defined for a (discontinuous) measurement-induced evolution~\cite{chruscinski2004geometric,berry1996,facchi1999}, in which case the discontinuous jumps of the trajectory are completed by shortest geodesics in quantum state space. 
Such measurement-induced geometric phases have been measured for strong~\cite{berry1996,do2019} and weak~\cite{cho2019} measurements, and can lead to topological transitions with respect to the measurement strength~\cite{gebhart2020topological,snizkho2021,wang2022,ferrer2022}. 

Measurement-induced geometric phases mirror earlier definitions of relative phases between optical beams~\cite{pancharatnam1956} or between quantum states~\cite{bargmann1964,avdoshkin2022}. 
In particular, the collective (Pancharatnam) phase 
$\phi_\mathrm{g}$ corresponding to the (ordered) tuple of $N$ quantum states $(\ket{\psi_1},\dots, \ket{\psi_N})$ is defined as~\cite{pancharatnam1956} 
\begin{equation}\label{eq:pan}
    \phi_\mathrm{g} = \arg \left[\bra{\psi_1}\psi_2\rangle\bra{\psi_2}\psi_3\rangle\cdots \bra{\psi_N}\psi_1\rangle\right].
\end{equation}
This phase is identical to the geometric phase induced by a series of projectors $\ket{\psi_N}\bra{\psi_N},\dots,\ket{\psi_1}\bra{\psi_1}$ (corresponding to a sequence of outcomes of a series of projective measurements) on the initial state $\ket{\psi_1}$ \cite{chruscinski2004geometric,berry1996,facchi1999}. 
Collective geometric phases naturally arise in the interference pattern of optical beams~\cite{pancharatnam1956} and of different partially distinguishable particles~\cite{kobayashi2010}, and have been analysed in theoretical~\cite{shchesnovich2018,minke2021,wu2022} and experimental~\cite{menssen2017,jones2020} studies. 

Observing nontrivial collective geometric phases requires the quantum states to be neither distinguishable nor perfectly indistinguishable: Highly indistinguishable states result in small geometric phases, while for highly distinguishable states the visibility of the phase vanishes. Thus, such phases appear to oppose a high amount of indistinguishability that is known to be useful for different quantum information processing tasks~\cite{killoran2014,morris2020}: 
After the pioneering work of Hong, Ou and Mandel~\cite{hong1987,zou1991}, the indistinguishability of quantum states has been used to develop various widely-used techniques, such as entanglement swapping~\cite{zukowski1993} or the creation of Greenberger-Horne-Zeilinger (GHZ) states~\cite{greenberger1990,bouwmeester1999,krenn2017}. 
In particular, the quantum phenomenon of Bell nonlocality~\cite{bell1964,bell1976,brunner2014} can be created from the interference of independent perfectly-indistinguishable particles~\cite{yurke1992,yurke1992b}, using a cyclic interferometer with tunable internal phase shifts. 
In this way, one can generate bipartite~\cite{yurke1992b}, multipartite~\cite{yurke1992}, and genuine multipartite~\cite{gebhart2021} nonlocality.

In this work, we show that Bell nonlocality can be observed in a fixed interferometric setup with single photon inputs. 
In particular, we consider a cyclic interferometer with fixed internal phase shifts and measurement stations. Instead, the parties of the Bell scenario choose different internal states of the photons that enter the interferometer. These states shape the measured interference pattern by means of a collective geometric phase between the partially distinguishable input photons, a means that we show is essential to create nonlocal correlations in any fixed interferometer with single photon inputs.
Creating nonlocal correlations in this way is complicated by the intrinsic trade off between the size of the geometric phases and their interferometric visibility.
We circumvent this obstacle by employing a dynamical version of the quantum Zeno effect~\cite{misra1977,peres1980,facchi1999,snizkho2020} such that, in the ideal noiseless case, the cyclic interferometer results in bipartite nonlocal correlations using a total of twelve photons, where the two parties each control the preparation of five input photons. 

 
\section{The cyclic interferometer}\label{sec:setup}

\begin{figure}[t]
		\center
		\includegraphics[width = 0.7\columnwidth]{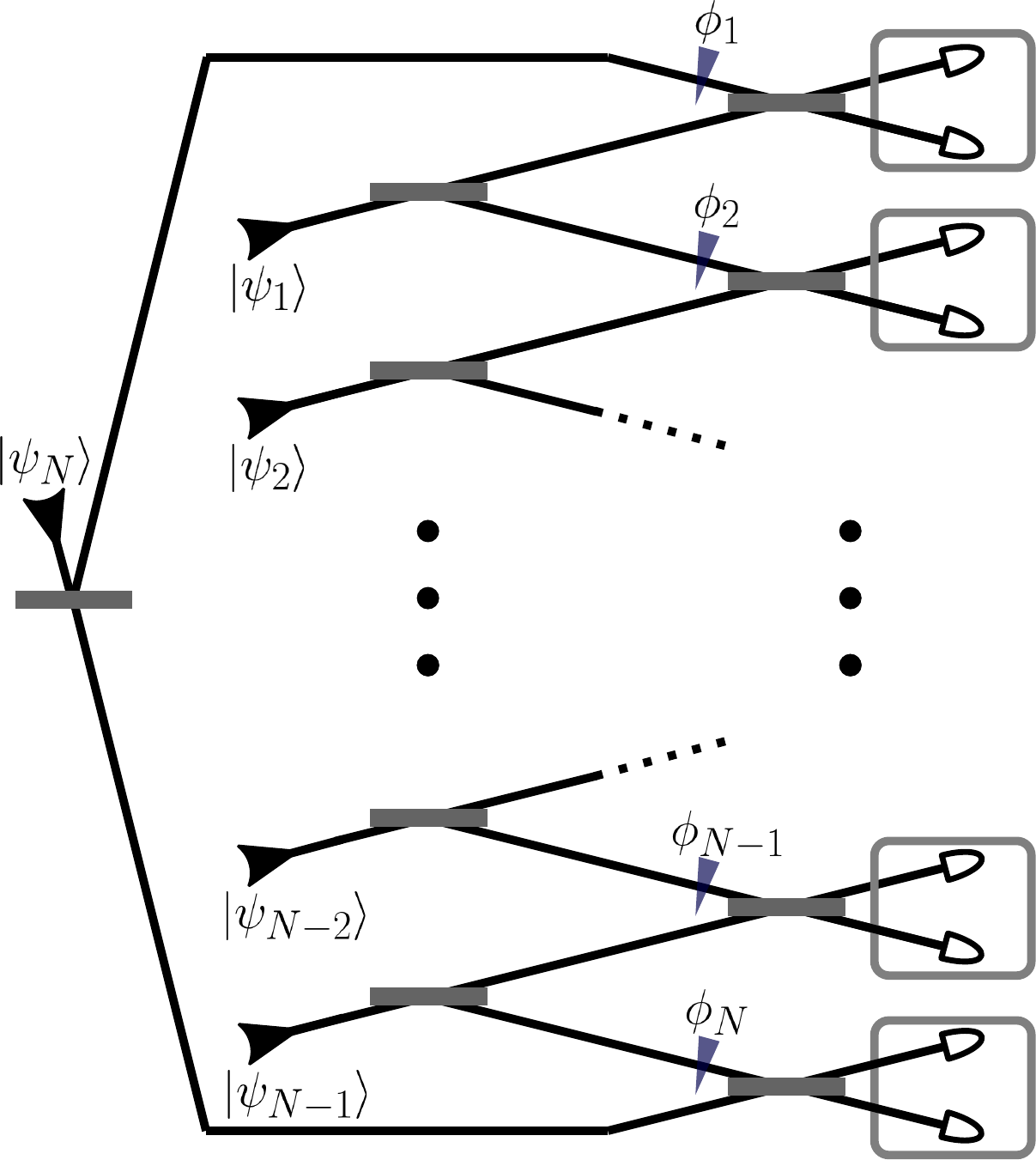} 
		\caption{
		    The $N$-photon cyclic interferometer~\cite{yurke1992}. The $m$th input photon, prepared in the internal state $\ket{\psi_m}$, is distributed by a balanced beam splitter between the $m$th and $(m+1)$th measurement stations. The $m$th measurement station consists of a phase shift $\phi_m$ on the upper incoming mode, a balanced beam splitter, and two numer-resolving detectors of the outgoing modes.
		}
		\label{fig:nyurke}
	\end{figure}

We consider a $N$-photon cyclic interferometer, in which $N$ independent single photons interfere in a circular optical circuit consisting of two layers of beam splitters (as introduced by Yurke and Stoler~\cite{yurke1992}), see Fig.~\ref{fig:nyurke}. Note that the results of this work can be derived for any bosonic or fermionic particles but we restrict ourselves to photons for simplicity. We assume that the photons are indistinguishable except for their state in an internal degree of freedom (e.g., their polarization, or their spatial or temporal profile). In this degree of freedom, the $m$th photon is prepared in the pure state $\ket{\psi_m}$. 
In the first layer, the $m$th photon enters a balanced beam splitter with outgoing modes directed to the $m$th  and $(m+1)$th measurement station (the $N$th photon is distributed between the $N$th and the first measurement station). Then, in the $m$th measurement station, the upper incoming mode experiences a phase shift $\phi_m$, after which the two modes interfere in a second balanced beam splitter. The outgoing modes of the beam splitters are then measured using number-resolving  detectors. 

In the following, we focus on the coincidence events for which every measurement station detects a single photon\footnote{Coincidence occurs with a probability $P=1/2^{N-1}$. This can be seen by counting the ($2^N$) different paths of the incoming photons after the first layer of beam splitters (each having the same amplitude because the beam splitters are balanced). Exactly two of these paths result in a single photon per party. For a detailed derivation see Appendix \ref{ap:YS_probs}.}. 
We label the outcome $o_m$ of the $m$th measurement station as $o_m=0$ ($o_m=1$) if the upper (lower) detector detects the photon. Furthermore, we denote the number of measurement stations that register $o_m=1$ as $k=\sum_m o_m$. If each photon is prepared in the same internal state (i.e., the incoming photons are perfectly indistinguishable), the probability that $k$ is even (odd) is given by $P(k=0\mod 2)=(1+(-1)^N\cos\phi)/2^N$ [$P(k=1\mod 2)=(1-(-1)^N\cos\phi)/2^N$]~\cite{yurke1992,pont2022}, where $\phi=\sum_m\phi_m$. 
If instead the $m$th photon is prepared in the internal state $\ket{\psi_m}$, the outcomes $\mathbf{o}=(o_1,o_2,\dots,o_N)\in\{0,1\}^N$ occur with probability 
\begin{equation} \label{eq:YS_probs}
P(\mathbf{o})= \frac{1}{2^{2N-1}}\left(1+(-1)^{N+k}\operatorname{Re}\left[Ve^{-i\phi}\right]\right),
\end{equation} 
where we defined the geometric factor
\begin{equation}\label{eq_Vpure}
V=\bra{\psi_1}\psi_2\rangle\bra{\psi_2}\psi_3\rangle\cdots \bra{\psi_N}\psi_1\rangle. 
\end{equation} 
For the derivation of Eq.~\eqref{eq:YS_probs}, see Appendix~\ref{ap:YS_probs}. The phase $\phi_\mathrm{g}=\arg [V]$ of the geometric factor is exactly the collective geometric phase, Eq.~\eqref{eq:pan}, corresponding to the states $(\ket{\psi_1},\dots,\ket{\psi_N})$ and acts as an offset to the phase $\phi$.
The absolute value $\left|V\right|$ can be seen as a decreasing many-photon interference visibility due to increasing distinguishability between the incoming photons.  We thus observe that, without changing the internal phase shifts $\phi_m$, we can affect the outcome probabilities by varying the internal states of the input photons. Finally, we note that also for the more realistic case that each photon is prepared in a mixed internal state, the outcome probabilities depend on a geometric phase factor between different mixed internal states, see Appendix~\ref{ap:mixed}.

\section{Creating nonlocality in a fixed interferometer}\label{sec:nonlocality}

We now fix the cyclic interferometer, i.e., we fix the internal phase shifts $\phi_m$, the beam splitters, and the number-resolving measurement stations. In the following, we demonstrate that one can generate nonlocal correlations by preparing the input photons in different internal states. This makes our approach distinct to earlier proposals where the internal phase shifts $\phi_m$ represent the measurement choices of the different parties~\cite{yurke1992,gebhart2021,gebhart2022coincidence}\footnote{It has been shown that a cyclic interferometer with $N$ independent (and perfectly indistinguishable) input photons and variable internal phase shifts $\phi_m$ can generate genuine $N$-partite nonlocality~\cite{yurke1992,gebhart2021,gebhart2022coincidence}. When postselecting on coincidence events, the outcome statistics correspond to a specific measurement of a $N$-particle GHZ state~\cite{greenberger1990,mermin1990,yurke1992}. Furthermore, in the noiseless case, the required postselection does not lead to any postselection loopholes~\cite{gebhart2021,gebhart2022coincidence}.}. 
Instead, we consider a scenario where the interferometer remains invariant and the only choices made by the parties are the internal states of the incoming photons\footnote{Note that, for the (original) definition of nonlocality that we consider here, a nontrivial influence of the parties' measurement choices on the outcome probabilities are necessary to generate nonlocality, see, e.g., Ref.~\cite{gebhart2022fairsampling} for a detailed argument. This is not true for the more recent definition of network nonlocality~\cite{navascues2020,supic2022}, in which nonlocal correlations can be observed in setups without measurement choices and, in particular, in a cyclic interferometer with fixed phase shifts $\phi_m$~\cite{abiuso2022}.}, representing a new approach to generate Bell nonlocality.

The influence of the internal states on the correlations is mediated exclusively by the geometric phases of Eq.~\eqref{eq:YS_probs}. 
Moreover, these geometric phases are crucial to create nonlocal correlations in any fixed interferometer, not just in the cyclic one that we consider in this work: In Appendix~\ref{ap:general_fock}, we prove that for an arbitrary fixed interferometer with independent Fock-state single-photon inputs and number-resolving measurements, nontrivial geometric factors, cf. Eq.~\eqref{eq_Vpure}, are necessary to generate nonlocality.

A first idea to generate nonlocality by varying the internal input states $\ket{\psi_m}$ is to identify the $m$th input photon as the measurement setting of the $m$th party. However, since each photon is distributed between two different measurement stations, the relativistic causal structure of this scenario differs from the one that is usually assumed in local hidden-variable models~\cite{brunner2014}, where each measurement setting has only one measurement outcome in its future light cone. A causal structure where each setting influences two outcomes requires a more sophisticated analysis, and possibly not even allows for quantum violations of Bell inequalities~\cite{chaves2017}\footnote{The setup that each setting can influence two outcomes (in a cyclic configuration) is considered in Ref.~\cite{chaves2017} for $N=3$ where it is argued that it is not clear whether one can find a Bell inequality that is violated by quantum correlations. Note that for $N=2$, a local hidden-variable model in which each setting influences two outcomes can simulate any bipartite statistics and, in particular, those generated by quantum mechanics.}. 
We thus restrict ourselves to a causal structure in which each party's input can only influence one measurement outcome. If each party inputs a single photon to the interferometer, it must therefore be associated with two measurement stations. In the bipartite Bell scenario, this corresponds to a four-photon cyclic interferometer.

\subsection{Four-photon cyclic interferometer}\label{sec:fourphotons}

\begin{figure}[t]
		\center
		\includegraphics[width = 0.9\columnwidth]{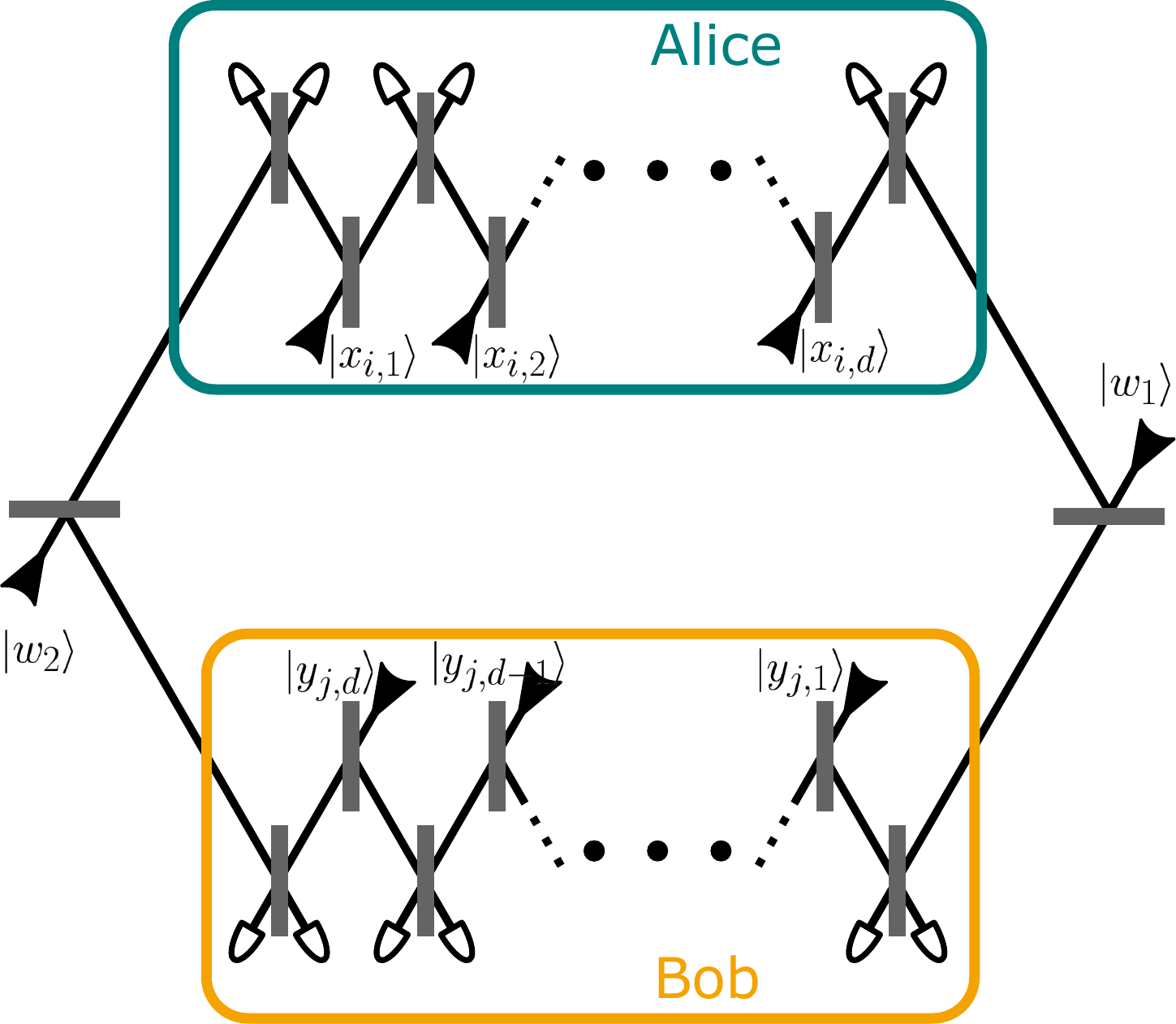} 
		\caption{
The rearrangement of the $(2d+2)$-photon cyclic interferometer (see Fig.~\ref{fig:nyurke}) into two parties, each having $(d+1)$ measurement stations and $d$ input photons. Alice's (Bob's) input photons are prepared in the initial states $(\ket{x_{i,1}},\dots,\ket{x_{i,d}})$ [ $(\ket{y_{j,1}},\dots,\ket{y_{j,d}})$], representing Alice's (Bob's) measurement setting. The two remaining input photons are prepared in the states $\ket{w_1}$ and $\ket{w_2}$. The setup can be seen as a bipartite Bell scenario where the two intermediate photons (prepared in the fixed states $\ket{w_1}$ and $\ket{w_2}$) constitute the shared common source, each measurement party has a binary measurement choice (among two different series of initial photon states for their $d$ local input photons), and each measurement party has a binary measurement output, given by a (local) post-processing of their $d+1$ local measurement outcomes).
		}
		\label{fig:multi}
	\end{figure}

We rearrange the four-photon setup such that each party includes one photon source and two measurement stations, see Fig.~\ref{fig:multi} with $d=1$. The two remaining input photons are prepared in the fixed internal states $\ket{w_1}$ and $\ket{w_2}$, and are distributed between the parties. The party Alice (Bob) prepares their photon in the internal state $\ket{x_{i,1}}$ ($\ket{y_{j,1}}$) depending on their measurement choice $i$ ($j$). As the interferometer and its internal phase shifts are fixed, we can assume to have calibrated the internal phase shifts such that $\phi=0$ in Eq.~\eqref{eq:YS_probs}. 


As above, we postselect the events for which each measurement station detects a single photon. Here, we must employ the fair-sampling assumption to close the detection loophole\footnote{In a standard Bell scenario, the fair-sampling assumption is not needed in an ideal loss-free experiment: Each party simply detects their particle and no events have to be postselected. Even in scenarios with a constant number of particles and a postselection of events in which each party detects a particle, such as the cyclic interferometer proposed by Yurke and Stoler, the fair-sampling assumption is unnecessary~\cite{blasiak2021,gebhart2021,gebhart2022coincidence}. This is because the postselection can be decided even when excluding any of the $N$ parties from the decision,  the so-called all-but-one principle~\cite{blasiak2021,gebhart2021}. Instead, in our setup, every party must be included in the postselection: Even if Alice detects one photon in each of her measurement stations, it is possible that Bob receives two photons in one of his measurement stations, and the event should be discarded as it does not depend on the $N$-photon collective phase. Therefore, we must assume fair sampling, i.e., that the measurement setting of each party does not influence the detection probability in possible local-realistic explanations~\cite{berry2010,gebhart2022fairsampling}.}.
The simplest Bell inequality in a bipartite scenario is the CHSH inequality~\cite{clauser1969} that assumes two measurement settings and two measurement outcomes per party. Since each party has two measurement stations, we merge the four possible outcomes into two: Alice's outcome is defined as $a=1$ ($a=-1$) whenever Alice measures $00$ or $11$ ($01$ or $10$). Bob merges his outcomes similarly to $b=1$ and $b=-1$. Using Eq.~\eqref{eq:YS_probs} (and renormalizing the probabilities after the postselection), we obtain 
\begin{equation}
P(a,b|i,j)=\frac{1}{4}(1+ab\operatorname{Re}\left[V_{ij}\right])
\end{equation}
with $V_{ij}=\bra{x_{i,1}}w_1\rangle\bra{w_1}y_{j,1}\rangle\bra{y_{j,1}}w_2\rangle \bra{w_2}x_{i,1}\rangle$.
We thus obtain the correlations 
\begin{equation}
    \langle A_iB_j\rangle = \sum_{a,b}abP(a,b|i,j) =  \operatorname{Re}\left[V_{ij}\right]. \label{eq:correlator_phase}
\end{equation}
To observe nonlocality, we must find states $\ket{x_{i,1}}$, $\ket{y_{j,1}}$, and $\ket{w_k}$  such that we violate the CHSH inequality, 
\begin{align}
    I_\mathrm{CHSH}&=\langle A_1B_1\rangle+\langle A_1B_2\rangle+\langle A_2B_1\rangle-\langle A_2B_2\rangle \notag \\ 
    &= \operatorname{Re}\left[V_{11}+V_{12}+V_{21}-V_{22}\right]\leq 2. \label{eq:CHSH}
\end{align}

If the internal Hilbert space of each photon is two-dimensional (e.g., if we use the photon's polarization as internal degree of freedom), we can numerically optimize Eq.~\eqref{eq:CHSH}, see Appendix~\ref{ap:fourphotons} for a detailed description. We find that there are no combinations of $\ket{x_{i,1}}$, $\ket{y_{j,1}}$, and $\ket{w_k}$ such that the inequality~\eqref{eq:CHSH} is violated. This is because to achieve a significant violation, the geometric phases (i.e., the solid angles) associated with the different $\ket{x_{i,1}}$, $\ket{y_{j,1}}$, and $\ket{w_k}$ must be large, reducing the visibility $\left|V_{ij}\right|$ because the photon states become more distinguishable. Thus, we must either consider higher dimensional internal Hilbert spaces\footnote{As a numerical check, we have generated $10^7$ random states $\ket{x_{i,1}}$, $\ket{y_{j,1}}$, and $\ket{w_k}$ for each $d=3,4,5$, all of which satisfied $I_\mathrm{CHSH}\leq 2$.}, or an $N$-photon cyclic interferometer with $N>4$ as we do in the following. 

\subsection{Multiphoton cyclic interferometer}

The obstacle of a decreasing multiphoton visibility for increasing geometric phases can be overcome by making use of the dynamical quantum Zeno effect~\cite{misra1977,peres1980,facchi1999,snizkho2020}. For a given number of states, we can insert additional states along the geodesics connecting the initial states, such that the geometric phase remains the same but the visibility increases. 
Therefore, we now consider a cyclic interferometer with $2d+2$ photon sources (and $2d+2$ measurement stations) and combine $d$ photon sources and $d+1$ measurement stations to form each party, see Fig.~\ref{fig:multi}. 
In the cyclic interferometer, the correlations are dependent on the collective geometric phase $\phi_\mathrm{g}$ corresponding to the internal states $(\ket{\psi_1},\dots,\ket{\psi_{2d+2}})$, see Eqs.~\eqref{eq:YS_probs} and \eqref{eq:correlator_phase}. We define two sequences of input states $(\ket{x_{i,1}},\dots,\ket{x_{i,d}})$ for Alice and two sequences of input states $(\ket{y_{j,1}},\dots,\ket{y_{j,d}})$ for Bob, representing the respective measurement settings $i,j\in\{1,2\}$. As in Sec.~\ref{sec:fourphotons}, we merge the $d+1$ different measurement outcomes of each party into two outcomes, defining $a=1$ ($b=1$) if Alice (Bob) observes an even number of $1$-outcomes, and $a=-1$ ($b=-1$) otherwise. Using Eq.~\eqref{eq:YS_probs}, one shows that each combination $(i,j)$ of measurement settings results the correlation $\left\langle A_i B_j \right\rangle=\operatorname{Re}[V_{ij}]=|V_{ij}|\cos(\phi_{\mathrm{g},ij})$, where $\phi_{\mathrm{g},ij}$ is the collective geometric phase associated with the input states of the measurement choices $(i,j)$. Note that the probability is again renormalized to events in which each measurement station detects a photon.

Let us emphasize in what sense the $(2d+2)$-photon interferometer of Fig.~\ref{fig:multi} can be thought of as a bipartite Bell scenario. The basic requirements in a bipartite Bell scenario are two measurement parties that each receive one part of a shared common source (assumed to be a classical hidden variable to derive Bell inequalities) and then perform local measurements (with independent choices of measurement settings) at spacelike separation~\cite{brunner2014}. In Fig.~\ref{fig:multi}, the two distributed photons in the fixed internal states $\ket{w_1}$ and $\ket{w_2}$ represent the shared source (and could in principle originate from a single lab). Each party implements a binary measurement choice by the preparation of two possible sets of local input photon states  ($(\ket{x_{i,1}},\dots,\ket{x_{i,d}})$ and $(\ket{y_{i,1}},\dots,\ket{y_{i,d}})$, respectively). Finally, each party produces a binary measurement outcome by post-processing their $(d+1)$ local measurement results. In principle, the two parties can be placed at a large distance, in which case the two distributed photons might be generated earlier than the party's local input photons to achieve coincidence. Therefore, all components of a bipartite Bell scenario are given.

To find a set of internal photon states that generates nonlocal correlations in this configuration, we build upon the ideas of Ref.~\cite{clauser1974}: If one has a family of measurement observables $A(\alpha)$ of Alice [$B(\beta)$ of Bob] that is described by a single parameter $\alpha$ ($\beta$), and it holds that the correlations between $A(\alpha)$ and $B(\beta)$ fulfill $\left\langle A(\alpha) B(\beta) \right\rangle=\cos(\alpha-\beta)$, then one can create nonlocal correlations by using measurement settings $\alpha_i$ and $\beta_j$ such that  
\begin{equation}\label{eq:CHalphas}
|\alpha_1-\beta_1|=|\alpha_1-\beta_2|=|\alpha_2-\beta_1|=|\alpha_2-\beta_2|/3.
\end{equation}
With these settings, one finds $I_\mathrm{CHSH} = 3\cos(\alpha_1-\beta_1) -\cos[3(\alpha_1-\beta_1)]$,
yielding the maximal violation of the CHSH inequality, $I_\mathrm{CHSH}=2\sqrt{2}$, for $|\alpha_1-\beta_1|=\pi/4$.
As motivated in Eq.~\eqref{eq:CHalphas}, we choose the sequences such that, in the Zeno limit $d\to\infty$, we have  $\phi_{\mathrm{g},11}=\phi_{\mathrm{g},12}=\phi_{\mathrm{g},21}=\phi_{\mathrm{g},22}/3$. This can be achieved by the trajectories depicted in Fig.~\ref{fig:trajectories} (shown for $d=3$), see Appendix~\ref{ap:geodesics} for the precise definition. In Fig.~\ref{fig:trajectories}, we have colored $\phi_{\mathrm{g},11}$ in blue, $\phi_{\mathrm{g},12}$ in red, and $\phi_{\mathrm{g},21}$ in yellow, and we note that $\phi_{\mathrm{g},22}=\phi_{\mathrm{g},11}+\phi_{\mathrm{g},12}+\phi_{\mathrm{g},21}$. 

\begin{figure}[t]
		\center
  \includegraphics[width = 0.98\columnwidth]{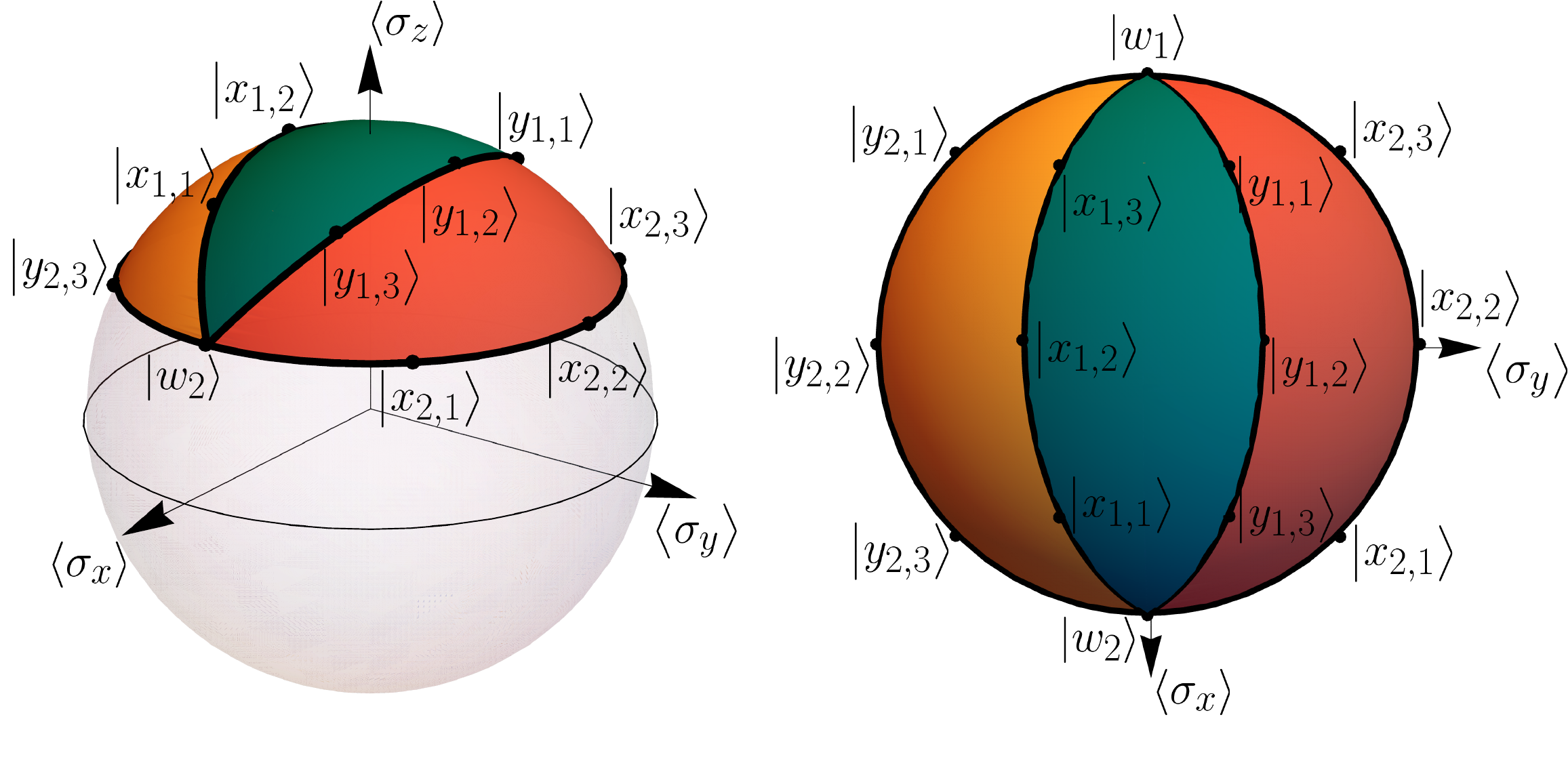} 
		\caption{
		The trajectories on the Bloch sphere that define the input photon states $(\ket{x_{i,1}},\dots,\ket{x_{i,d}})$ [$(\ket{y_{j,1}},\dots,\ket{y_{j,d}})$] corresponding to the measurement setting $i=1,2$ ($j=1,2$) of Alice (Bob), seen from the side (left panel) and from the top (right panel). The input states are sketched for $d=3$ input photons per party. The collective geometric phase $\phi_{\mathrm{g},ij}$ is proportional to the solid angle enclosed by the trajectories of measurement settings $(i,j)$. The geometric phase $\phi_{\mathrm{g},11}$ is colored in blue, $\phi_{\mathrm{g},12}$ in red, $\phi_{\mathrm{g},21}$ in yellow. The latitude of the horizontal trajectories is given by $\cos(\theta)$. $\left\langle \sigma_k \right\rangle$ is the expectation value of the  Pauli matrices, $k=x,y,z$.
		}
		\label{fig:trajectories}
	\end{figure}

\begin{figure}[t]
		\center
		\includegraphics[width = 0.99\columnwidth]{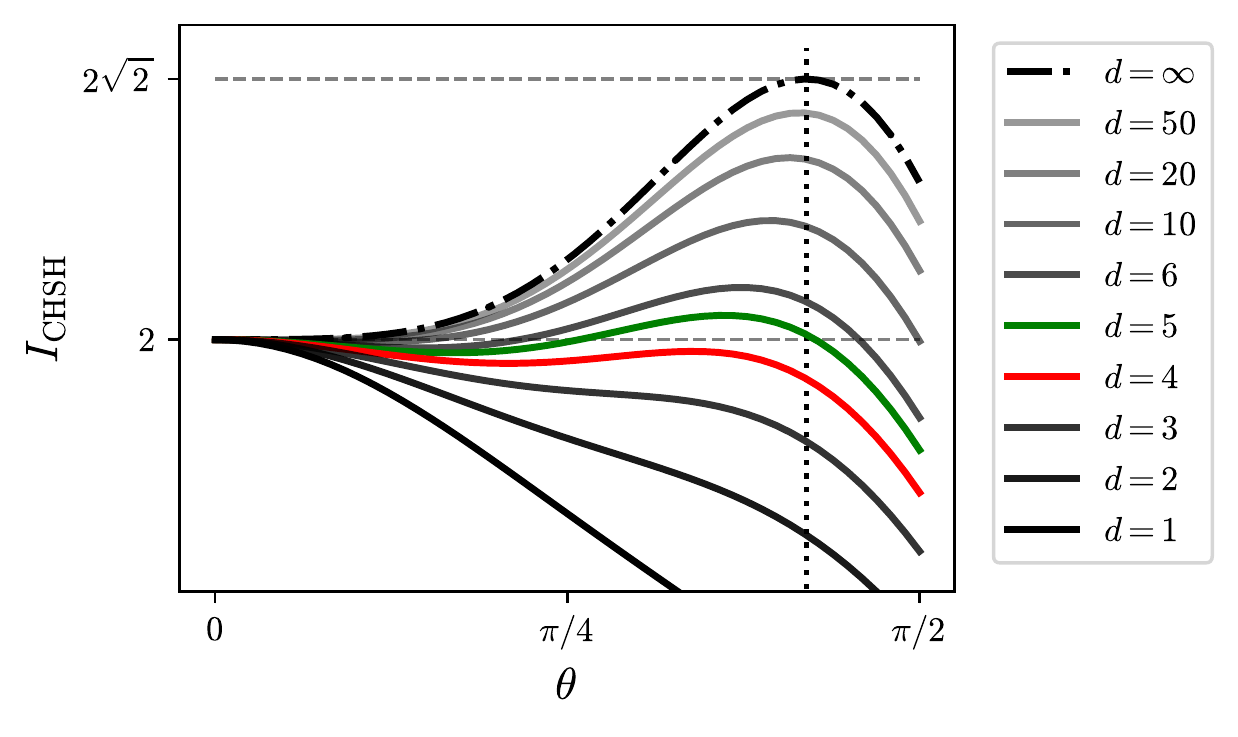} 
		\caption{$I_\mathrm{CHSH}$ as a function of the latitude $\theta$ of the horizontal trajectories in Fig.~\ref{fig:trajectories}, for different values of $d$, the number of input photons per party. The Zeno limit $d\to \infty$ is depicted as a dash-dotted line. For $d\geq5$ ($d=5$ in green), it is possible to violate the CHSH inequality, Eq.~\eqref{eq:CHSH}. The total number of photons is $N=2d+2$. The horizontal lines correspond to the CHSH inequality $I_\mathrm{CHSH}\leq2$, and the quantum bound ($I_\mathrm{CHSH}\leq2\sqrt{2}$). The vertical dotted line is the latitude $\theta=\arccos{1/4}$ for which the maximal violation of the CHSH inequality occurs. 
		}
		\label{fig:finite_d}
	\end{figure}

In the limit of a large number of photons, $d\to\infty$, the quantum Zeno effects results in $|V_{ij}|\to 1$.
In Fig.~\ref{fig:finite_d}, we show $I_\mathrm{CHSH}$ for $d=500$ as a function of the latitude $\theta$ of the horizontal trajectory in Fig.~\ref{fig:trajectories} (i.e., $\theta=\arccos{\operatorname{tr}[\sigma_z \ketbra{w_2}]}$). 
For increasing latitude $\theta$, the solid angles enclosed by the trajectories (and thus the geometric phases) increase, resulting in an increasing value of $I_\mathrm{CHSH}$. The maximal quantum violation $I_\mathrm{CHSH}=2\sqrt{2}$ is reached for $\theta=\arccos{1/4}$ (vertical dotted line in Fig.~\ref{fig:finite_d}). This is consistent with the maximal value $|\alpha_1-\beta_1|=\pi/4$ in Eq.~\eqref{eq:CHalphas} because the geometric phase corresponding to the latitude $\theta$ is given by $ \phi_{\mathrm{g},22} = \pi (1-\cos\theta)$~\cite{berry1996}, and for $\theta=\arccos{1/4}$, we have $\phi_{\mathrm{g},11}=\phi_{\mathrm{g},22}/3=\pi (1-\cos\theta)/3=\pi/4$.

We can thus violate the CHSH inequality up to the maximal quantum violation in the Zeno limit $d\to\infty$, and we saw in Sec.~\ref{sec:fourphotons} that cannot violate it for $d=1$. Since the cyclic interferometer with independent input photons is increasingly difficult to implement for larger $d$, it is natural to ask what the minimal value of $d$ is that suffices to observe nonlocality. 
In Fig.~\ref{fig:finite_d}, we show the values $I_\mathrm{CHSH}$ as a function of the latitude $\theta$ for several values of $d$. Note that the $d$ different states for each measurement setting were chosen on the respective trajectories such that two adjacent states have the same (absolute) overlap, see Appendix~\ref{ap:geodesics}. For our choice of input photon states, we thus observe a violation of the CHSH inequality for $d\geq 5$ (green curve), corresponding to a cyclic interferometer with a total of $12$ input photons. 

We would like to note that, in an analogous way, one could also create genuine multipartite nonlocality. Here, the $N$ input photons must be divided among at least three parties, and the respective states should be chosen such that a corresponding multipartite Bell inequality is violated~\cite{svetlichny1987}. However, due to the larger number of parties, we expect the required number of input photons to be much larger than in the bipartite case.

\section{Conclusions}\label{sec:conclusions}
In this work, we have employed a 
static cyclic $N$-photon interferometer to generate Bell nonlocality from independent partially-distinguishable photon sources. 
Importantly, the interferometer (i.e., its internal phase shifts and the subsequent measurements) is fixed and the parties' measurement settings are instead provided by the choice of the internal states of the input photons, representing a new approach to generate nonlocality. 
The internal states shape the outcome probabilities through a collective $N$-photon geometric phase that can be used to (and is necessary to) generate nonlocal correlations in a fixed interferometer.
The observation of nonlocal correlations in this way is hindered by an intrinsic trade off between the $N$-photon interference visibility (that increases for less distinguishable internal states) and the collective $N$-photon geometric phases (that increases for more distinguishable internal states). 
We have overcome this obstacle by using the dynamical quantum Zeno effect, yielding nonlocal correlations for an interferometer with $N\geq 12$ independent single-photon inputs. 

\section*{Acknowledgments}
The author would like to thank A.~Smerzi and M.~Gessner for helpful discussions. This work was supported by the European Commission through the H2020 QuantERA ERA-NET Cofund in Quantum Technologies project “MENTA”.

\appendix
\section*{Appendix}

\section{Cyclic interferometer with partially distinguishable photons}\label{ap:YS_probs}
In this appendix, we calculate the outcome probabilities of the $N$-photon cyclic interferometer with partially distinguishable internal states of the input photons, see Fig.~\ref{fig:nyurke}. Let the $m$th incoming photon be prepared in the internal state $\ket{\psi_m}$. We write $\ket{\psi_m}=a^\dagger_m(\psi_m)\ket{0}$, where we index the creation/annihilation operators by $m$ indicating the $m$th input mode, and $\ket{0}$ is the vacuum mode. The first layer of beam splitters is described by the transformations \begin{equation} \label{eq:firstBS}
\begin{pmatrix} b^\dagger_{m,1} (\psi) \\ b^\dagger_{m+1,0} (\psi) \end{pmatrix} 
= \frac{1}{\sqrt{2}}\begin{pmatrix} 1 & -i \\ -i & 1 \end{pmatrix}
\begin{pmatrix} a^\dagger_m (\psi) \\ \tilde{a}^\dagger_m (\psi) \end{pmatrix},
\end{equation}
where $\tilde{a}_m$ corresponds to the second input port of the beam splitters (that is always prepared in the vacuum state), and the labels of the outgoing modes ($b$) are chosen such that the $m$th input photon is divided between the lower ($1$-mode) incoming arm of the $m$th measurement station and the upper ($0$-mode) incoming arm of the $(m+1)$th measurement station, see Fig.~\ref{fig:nyurke}. Note that the $N$th input photon is divided between the $N$th and the first measurement stations, so in the following, the mode index should be understood $\mod N$. Next, each upper incoming arm obtains a phase shift, $c^\dagger_{m,0} (\psi) = e^{-i\phi_m} b^\dagger_{m,0} (\psi)$, while the lower incoming arm is unchanged, $c^\dagger_{m,1} (\psi) = b^\dagger_{m,1} (\psi)$. The second layer of beam splitters finally produces the outgoing (detection) modes 
\begin{equation}
\begin{pmatrix} d^\dagger_{m,0} (\psi) \\ d^\dagger_{m,1} (\psi) \end{pmatrix} 
= \frac{1}{\sqrt{2}}\begin{pmatrix} 1 & -i \\ -i & 1 \end{pmatrix}
\begin{pmatrix} c^\dagger_{m,1} (\psi) \\ c^\dagger_{m,0} (\psi) \end{pmatrix},
\end{equation}

\noindent where the upper (lower) detection mode is labeled by $0$ ($1$). Using these transformations, one finds that
\begin{align}\label{eq:allBSs}
    a^\dagger_m(\psi_m) &= \frac{1}{2}\big[d^\dagger_{m,0}(\psi_m)+id^\dagger_{m,1}(\psi_m) \notag \\&\quad \quad \quad +e^{i\phi_{m+1}}\left(id^\dagger_{m+1,1}(\psi_m)-d^\dagger_{m+1,0}(\psi_m)\right)\big].
\end{align}

Let us first calculate the probability $P(\mathbf{0})$ of all measurement stations detecting a photon in the upper ($0$-mode) arm. This outcome is associated with the projector $\operatorname{proj}_\mathbf{0}$ which is given by 
\begin{equation}\label{eq:proj0}
    \operatorname{proj}_\mathbf{0} = \bigotimes_m \left[\sum_k d^\dagger_{m,0}(\xi_k)\ket{0}\bra{0}d_{m,0}(\xi_k)\right], 
\end{equation}
where $\{\ket{\xi_k}\}_k$ is an arbitrary basis of the internal Hilbert space. We must sum over the internal Hilbert space because the detection does not depend on the internal state of the photon. We need to calculate $P(\mathbf{0})=\bra{\Psi}\operatorname{proj}_\mathbf{0} \ket{\Psi}$, where $\ket{\Psi}=\bigotimes_m a^\dagger_m(\psi_m) \ket{0}$ is the initial state. The only terms of $\ket{\Psi}$ that contribute to $P(\mathbf{0})$ are the ones where each $0$-mode has one photon, i.e., the terms including $\bigotimes_m d^\dagger_{m,0}(\psi_m)$ and the terms including $\bigotimes_m d^\dagger_{m+1,0}(\psi_m)$, such that we are left with 
\begin{widetext}
  \begin{align}
    P(\mathbf{0}) &= \frac{1}{2^{2N}}\left\lVert \left[\bigotimes_m d^\dagger_{m,0}(\psi_m) + (-1)^N e^{i\phi}\bigotimes_m d^\dagger_{m+1,0}(\psi_m) \right]\ket{0} \right\rVert ^2 \\ 
    &= \frac{1}{2^{2N}} \left\{2 + 2(-1)^N \operatorname{Re}\left[e^{i\phi}\bra{0}\bigotimes_m d_{m,0}(\psi_m)\bigotimes_m d^\dagger_{m+1,0}(\psi_m)\ket{0}\right]\right\},
\end{align}  
\end{widetext}
where we have used $\phi=\sum_m\phi_m$. 
By writing $\bigotimes_m d^\dagger_{m+1,0}(\psi_m)\ket{0}=\ket{\psi_N}\otimes\ket{\psi_1}\otimes\dots\otimes \ket{\psi_{N-1}}$ and $\bra{0}\bigotimes_m d_{m,0}(\psi_m)=\bra{\psi_1}\otimes\dots\otimes \bra{\psi_{N}}$, we find 
\begin{equation}
   P(\mathbf{0}) = \frac{1}{2^{{2N}-1}} \left(1+(-1)^N \operatorname{Re}\left[V e^{-i\phi}\right]\right)
\end{equation}
with $V=\bra{\psi_1}\psi_2\rangle\bra{\psi_2}\psi_3\rangle\cdots \bra{\psi_N}\psi_1\rangle$, see Eq.~\eqref{eq_Vpure}. 

To obtain $P(\mathbf{o})$ for $\mathbf{o}\neq \mathbf{0}$, note that for each $1$-entry in $\mathbf{o}$, one of the two terms of $\ket{\Psi}$ that contribute to $P(\mathbf{o})$ is multiplied by $i$ (due to reflection instead of transmission), while the other term is multiplied by $-i$ (due to a transmission instead of reflection), resulting in a additional relative phase of $-1$ between the two terms. Thus, we find Eq.~\eqref{eq:YS_probs},
\begin{equation}P(\mathbf{o})= \frac{1}{2^{2N-1}}\left(1+(-1)^{N+k}\operatorname{Re}\left[Ve^{i\phi}\right]\right),
\end{equation}
where $k=\sum_m o_m$ is the number of measurement stations detecting $o_m=1$. We emphasize that this equation only holds when each measurement station detects one photon. Analogous to the results of Ref.~\cite{yurke1992}, all other events have constant probability, i.e., the corresponding probabilities depend neither on the phase shifts $\phi_m$ nor on the internal states $\ket{\psi_m}$.

\section{Mixed internal states}\label{ap:mixed}

In this section, we analyze what happens in the (more realistic) case where each party can only insert mixed states $\rho_i$ instead of pure states $\ket{\psi_i}$ to the interferometer. We first rewrite Eq.~\eqref{eq_Vpure} for pure input states $\ket{\psi_i}$ as $V = \operatorname{tr}\left[\ket{\psi_1}\bra{\psi_1}\cdots \ket{\psi_N}\bra{\psi_N}\right]$, where $\ket{\psi_i}\bra{\psi_i}$ is the density matrix corresponding to the pure state $\ket{\psi_i}$. If we use the mixed states $\rho_i$ instead, it follows from linearity and Eq.~\eqref{eq:YS_probs} 
that the interference depends on the geometric factor
\begin{equation}\label{eq:Vmixed}
    V_\mathrm{mixed}= \operatorname{tr}\left[\rho_1\rho_2\cdots\rho_N\right].
\end{equation}

This geometric factor can be compared to different (non-equivalent) generalizations of the geometric phase for mixed states~\cite{uhlmann1986,dabrowski1989,sjoqvist2000,ericsson2003,peixoto2003,chaturvedi2004}. First, by decomposing the internal states as $\rho_i=\sum_{k}c_{k}^{(i)}\ket{c_{k}^{(i)}}\bra{c_{k}^{(i)}}$, we see that $V_\mathrm{mixed}$ can be interpreted as an average of different pure-state geometric factors $V$, 
\begin{equation}
  V_\mathrm{mixed} =   \sum_{k, \dots ,m}c_{k}^{(1)}\cdots c_{m}^{(N)}V\left(\ket{c_{k}^{(1)}},\dots,\ket{c_{m}^{(N)}}\right),
\end{equation}
where $V\left(\ket{c_{k}^{(1)}},\dots,\ket{c_{m}^{(N)}}\right)$ is the collective phase, Eq.~\eqref{eq:pan}, corresponding to the states $(\ket{c_{k}^{(1)}},\dots,\ket{c_{m}^{(N)}})$.

Second, in Ref.~\cite{sjoqvist2000}, the geometric factor $V_\mathrm{sv}$ acquired by a mixed initial state $\rho_0=\sum_k c_k \ket{c_k}\bra{c_k}$ in a unitary evolution $U$ is defined as 
\begin{equation}
  V_\mathrm{sv}(\rho_0,U)= \operatorname{tr}\left[U\rho_0\right]=\sum_k c_k \left\langle c_k | U | c_k\right\rangle.
\end{equation}
i.e., as a weighted average over the geometric factors $\left\langle c_k | U | c_k\right\rangle$ associated with the pure states in the decomposition of $\rho$. This definition can be directly extended to a projection-induced evolution $P_{\ket{\psi_2},\dots,\ket{\psi_N}}$ 
that maps $\ket{\psi} \mapsto \ket{\psi_2}\bra{\psi_2}\psi_3\rangle\cdots \bra{\psi_N}\psi\rangle$, see Eq.~\eqref{eq:pan}: For such an evolution, we define the mixed-state geometric factor $V_\mathrm{sv}\left(\rho_0,P_{\ket{\psi_2},\dots,\ket{\psi_N}}\right)= \sum_k c_k V_\mathrm{proj}\left(\ket{c_k},P_{\ket{\psi_2},\dots,\ket{\psi_N}}\right)$, 
where $V_\mathrm{proj}$ is the geometric factor, Eq.~\eqref{eq:pan}, defined for a projection-induced evolution~\cite{chruscinski2004geometric,berry1996,facchi1999}.
In our setup, by expanding each internal state $\rho_i$ as above, we see that the factor $V_\mathrm{mixed}$ is equivalent to a weighted average of the geometric factors $V_\mathrm{proj}$ of different projection-induced evolutions of the initial state $\rho_1$, 
 \begin{equation}
V_\mathrm{mixed} = \sum_{l, \dots ,m}c_{l}^{(2)}\cdots c_{m}^{(N)}V_\mathrm{sv}\left(\rho_1,P_{\ket{\psi_{l}^{(2)}},\dots,\ket{\psi_{m}^{(N)}}}\right).
 \end{equation}
Note that, by adjusting the ensemble of projections accordingly, any of the input states $\rho_i$ can be treated as the initial state.

Finally, each internal state $\rho_i$ 
can also be seen as an operator of a generalized (non-projective) measurement~\cite{breuer2002theory}. In this description, the geometric factor $V_\mathrm{mixed}$ corresponds to the geometric factor defined for a series of Kraus operators defined for a completely-positive map~\cite{ericsson2003,peixoto2003}. 

\section{Fixed interferometer with single-port single-photon inputs}\label{ap:general_fock}
Here, we show why, in a general fixed interferometer with $N$ independent single-photon input states that are inserted in fixed modes, nontrivial geometric factors [cf. Eq.~\eqref{eq_Vpure} for pure states and Eq.~\eqref{eq:Vmixed} for mixed states] between the internal input states are necessary to generate nonlocal correlations. In particular, we show that in any such interferometer, the only possible influence of the internal states of the input photons on the outcome probabilities is caused by geometric factors (geometric phases and their visibilities). Furthermore, for the cyclic interferometer considered in this work, cf. Fig.~\ref{fig:multi}, these geometric factors must include nontrivial geometric phases.  We also note that if we relax the assumption that we use single-port photons that are inserted into the fixed interferometer, geometric factors between the internal states are not necessary to generate nonlocality anymore:
It can be shown that one can create nonlocal correlations in a fixed interferometer by inserting the single photons in measurement-setting-dependent superpositions of the input ports, where all photons are prepared in the same internal state (thus corresponding to a trivial geometric factor). However, these superpositions must be created by precedent interferometers that must be variable, so strictly speaking the interferometer is not fixed in this case.

The importance of geometric phase factors follows directly from the analysis performed by Shchesnovich~\cite{shchesnovich2014,shchesnovich2015,shchesnovich2018}, which we will summarize in the following. Consider a general interferometer with $K$ input ports and $K$ output ports. The interferometer is described by a unitary $U$ relating the input modes $a_k^\dagger$ to the output modes $d_l^\dagger$, $a_k^\dagger = \sum_{l=1}^{K} U_{k,l} d_l^\dagger$. Note that here we label incoming and outgoing modes differently to Appendix~\ref{ap:YS_probs}. Since the interferometer is arbitrary, we can assume that the $i$th input photon, prepared in the internal state $\rho_i$, is inserted in the $i$th input port of the interferometer. In Refs.~\cite{shchesnovich2014,shchesnovich2015}, it is shown that the probability $P(\mathbf{l})$ of measuring the photons in the outgoing modes $\mathbf{l}=(l_1,\dots,l_N)$ (detecting multiple photons in a single output port is represented by repeating the output label correspondingly in $\mathbf{l}$) is given by 
\begin{equation}
    P(\mathbf{l}) = \frac{1}{M(\mathbf{l})} \sum_{\tau,\sigma \in S_N} J(\tau^{-1}\sigma)\prod_{k=1}^N U^*_{k,l_{\tau(k)}}U_{k,l_{\sigma(k)}}.\label{eq:shches}
\end{equation}
Here, $M(\mathbf{l})=\prod_{i=1}^{K}\left[\left(\sum_{j=1}^N\delta_{l_j,i}\right)!\right]$ is a combinatorial factor accounting for multiphoton detection in a single output port, $\tau$ and $\sigma$ are permutations (i.e., elements of the symmetric group of degree $N$, $S_N$), and the function $J$ describes the distinguishability of the internal states $\rho_i$: If we denote the set of disjoint cycles generating $\sigma$ as $cyc(\sigma)$, the function $J(\sigma)$ is given by~\cite{shchesnovich2014,shchesnovich2015,shchesnovich2018}
\begin{equation}
    J(\sigma)=\prod_{\gamma\in cyc(\sigma)}  \operatorname{tr}\left[\rho_{k_{r_\gamma}}\cdots\rho_{k_2}\rho_{k_1}\right], 
\end{equation}
where $\gamma=(k_1\,k_2\,\dots\,k_{r_\gamma})$ denotes the different cycles in $cyc(\sigma)$. Note that we assume perfect number-resolving detectors and no losses in the interferometer. 

Since we consider a fixed interferometer, the terms $U_{k,l}$ are constant, and we deduce from Eq.~\eqref{eq:shches} that the only influence of the input states $\rho_i$ on the output probabilities $P(\mathbf{l})$ is due to terms of the form $\operatorname{tr}\left[\rho_{k_{r}}\cdots\rho_{k_2}\rho_{k_1}\right]$. These terms correspond exactly to the geometric phases factors (i.e., geometric phases and their visibilities) of different ensembles of internal states, see Eq.~\eqref{eq_Vpure} and Eq.~\eqref{eq:Vmixed}. In particular, for pure internal states, these terms include the collective phase of Eq.~\eqref{eq:pan}. We note however that the possible geometric factors also include simple overlaps, such as $\operatorname{tr}\left[\rho_{i}\rho_{j}\right]$, that have a trivial geometric phase. These overlaps play the crucial role, e.g., in the Hong--Ou--Mandel effect~\cite{hong1987}.

We thus see that, if we use a fixed interferometer with an input consisting of independent single-port photons, the only influence of the internal states $\rho_i$ on the output probability is mediated by the geometric factors between different combinations of the $\rho_i$. Therefore, if these internal states correspond to the measurement settings in a Bell scenario, nontrivial geometric factors are necessary to create nonlocal correlations. Can we also prove the stronger claim that nontrivial geometric phases are necessary for nonlocality in a fixed interferometer with single-port single-photon inputs?

For the cyclic interferometer considered in this work, this is true: If the collective phase of the cyclic interferometer, Eq.~\eqref{eq_Vpure}, is positive for all different settings $(i,j)$, one can write 
\begin{equation}
    V_{ij}=r_{i}s_j
\end{equation}
with $r_i,r_j>0$. We then have $P(a,b|i,j)=(1+ab r_is_j)/4$, and one easily checks that this correlation cannot violate any CHSH inequality, and hence admits a local hidden variable model~\cite{fine1982}. Even if we do not merge the measurement outcomes of each party into only two possibilities, the local hidden variable model still holds because the probabilities of all outcomes $\mathbf{o}$ contributing to the merged outcomes $(a,b)$ are equal, cf. Eq.~\eqref{eq:YS_probs}. Therefore, if the collective phases of the internal states are trivial, the induced correlations can be described by a local hidden variable model, and thus do not show nonlocality. We were not able to prove this result for generating nonlocality in a general fixed interferometer.

\section{Four-photon interferometer cannot create nonlocal correlations}\label{ap:fourphotons}
Here, we briefly sketch how to numerically optimize Eq.~\eqref{eq:CHSH}, 
\begin{equation*}
    I_\mathrm{CHSH}
    = \operatorname{Re}\left[V_{11}+V_{12}+V_{21}-V_{22}\right], 
\end{equation*}
where $V_{ij}=\bra{x_{i,1}}w_1\rangle\bra{w_1}y_{j,1}\rangle\bra{y_{j,1}}w_2\rangle \bra{w_2}x_{i,1}\rangle$, if the internal Hilbert space is two dimensional.
We first write $\ket{x_{i,1}}\bra{x_{i,1}}=(\mathbb{I}+\mathbf{x}_i\cdot \boldsymbol{\sigma})/2$ and similarly for the other projectors, where $\mathbb{I}$ is the identity, $\boldsymbol{\sigma}$ are the Pauli matrices, and $\mathbf{x}_i$ is the corresponding Bloch vector. We can then use the identity $(\mathbf{r}\cdot \boldsymbol{\sigma})(\mathbf{s}\cdot \boldsymbol{\sigma})=(\mathbf{r}\cdot\mathbf{s})\mathbb{I}+i(\mathbf{r}\times\mathbf{s})\boldsymbol{\sigma}$ and obtain, after some simplifications,
\begin{widetext}
   \begin{align}
    I_\mathrm{CHSH}
    &= \frac{1}{16}\operatorname{Re}\big\{ \operatorname{Tr}\big[(\mathbb{I}+\mathbf{x}_1\cdot\boldsymbol{\sigma})(\mathbb{I}+\mathbf{w}_1\cdot\boldsymbol{\sigma})(2\mathbb{I}+(\mathbf{y}_1+\mathbf{y}_2)\cdot\boldsymbol{\sigma})(\mathbb{I}+\mathbf{w}_2\cdot\boldsymbol{\sigma}) \\ &\quad \quad \quad \quad\quad \quad+ (\mathbb{I}+\mathbf{x}_2\cdot\boldsymbol{\sigma})(\mathbb{I}+\mathbf{w}_1\cdot\boldsymbol{\sigma})((\mathbf{y}_1-\mathbf{y}_2)\cdot\boldsymbol{\sigma})(\mathbb{I}+\mathbf{w}_2\cdot\boldsymbol{\sigma})\big]\big\} \\ 
    &= \frac{1}{8}\big[ (1+\mathbf{x}_1\cdot \mathbf{w}_1)(2+(\mathbf{q}_0+\mathbf{q}_1)\cdot  \mathbf{w}_2) + (\mathbf{x}_1+ \mathbf{w}_1)\cdot(\mathbf{q}_0+\mathbf{q}_1 +2\mathbf{w}_2) -2 (\mathbf{x}_1\times \mathbf{w}_1)\cdot ((\mathbf{q}_0+\mathbf{q}_1)\times\mathbf{w}_2) \\
    &\quad \quad \quad \quad\quad \quad + (1+\mathbf{x}_2\cdot \mathbf{w}_1)((\mathbf{q}_0-\mathbf{q}_1)\cdot  \mathbf{w}_2) + (\mathbf{x}_2+ \mathbf{w}_1)\cdot(\mathbf{q}_0-\mathbf{q}_1) - (\mathbf{x}_2\times \mathbf{w}_1)\cdot ((\mathbf{q}_0-\mathbf{q}_1)\times\mathbf{w}_2) \big] 
\end{align} 
\end{widetext}
This expression can be numerically optimized over the Bloch vectors $\mathbf{x}_1$, $\mathbf{x}_2$, $\mathbf{w}_1$, $\mathbf{w}_2$, $\mathbf{y}_1$, and $\mathbf{y}_2$, e.g., by using spherical coordinates for each Bloch vector. The numerical optimization yields $I_\mathrm{CHSH}\leq 2$.

\section{Internal photon states to create nonlocality}\label{ap:geodesics}
Here, we describe how to find the trajectories on the Bloch sphere that correspond to the different measurement settings of the parties, in such a way that the geometric phases (approximately) fulfill $\phi_{\mathrm{g},11}=\phi_{\mathrm{g},12}=\phi_{\mathrm{g},21}=\phi_{\mathrm{g},22}/3$. Since $\phi_{\mathrm{g},22}$ corresponds to the largest solid angle, we choose it to be enclosed by a horizontal section of the Bloch sphere at latitude $\theta$, optimizing the solid angle with respect to a given visibility (which is determined by the absolute value of the overlaps between adjacent states in the trajectories). Thus, we define the trajectories $\mathbf{x}_2(t)$ and $\mathbf{y}_2(t)$ corresponding to Alice's setting $x_2$ and Bob's setting $y_2$ as 
\begin{equation}
\mathbf{x}_2(t) = \begin{pmatrix}
\sin \theta \cos t \\ \sin \theta \sin t \\ \cos \theta 
\end{pmatrix}  \quad \mathrm{and}  \quad \mathbf{y}_2(t) = \begin{pmatrix}
-\sin \theta \cos t \\ -\sin \theta \sin t \\ \cos \theta 
\end{pmatrix},
\end{equation}
with $t\in [0,\pi]$. The fixed states $\ket{w_1}$ and $\ket{w_2}$ correspond to the Bloch vectors $\mathbf{w}_1=(-\sin \theta,0,\cos \theta)^T$ and $\mathbf{w}_2=(\sin \theta,0,\cos \theta)^T$, connecting the two trajectories.

To define the trajectory corresponding to $x_1$, we want the trajectory to possesses the same starting and end points as $\mathbf{x}_2$, $\mathbf{x}_1(0)=\mathbf{x}_2(0)=\mathbf{w}_2$ and $\mathbf{x}_1(\pi)=\mathbf{x}_2(\pi)=\mathbf{w}_1$, but that it passes through $\mathbf{x}_1(\pi/2)=(0,-\sin \theta/3,\cos \theta/3)^T$ instead of $\mathbf{x}_2(\pi/2)=(0,\sin \theta,\cos \theta)^T$. We want to use this trajectory because, as seen in Fig.~\ref{fig:trajectories}, the points $\mathbf{x}_i(\pi/2)$ and $\mathbf{y}_j(\pi/2)$ are equidistant (in geodesic distance), except for a three times larger distance for $i=j=2$. We note that, for $\theta\neq \pi/2$, the corresponding geometric phases do not precisely fulfill the condition $\phi_{\mathrm{g},11}=\phi_{\mathrm{g},12}=\phi_{\mathrm{g},21}=\phi_{\mathrm{g},22}/3$ but they approximately fulfill it such that they are sufficient for our purposes. 

We now calculate the trajectory $\mathbf{x}_1(t)$. The plane that is defined by the three points $\mathbf{x}_1(t)$ for $t=0,\pi/2,\pi$ correspond to the solutions $(x,y,z)\in\mathbb{R}^3$ of the equation 
\begin{equation}
    [\cos(\theta/3)-\cos(\theta)]y+\sin(\theta/3)z -\cos(\theta)\sin(\theta/3)=0.
\end{equation}
We can eliminate $z$,
\begin{equation}
z=-2\sin(2\theta/3)y+\cos(\theta),
\end{equation}
where we have used that $(\cos(\theta)-\cos(\theta/3))/\sin(\theta/3)=-2\sin(2\theta/3)$. Inserting this expression for $z$ in the spherical equation $x^2+y^2+z^2=1$, we find (after completing the square for variable $y$) the equation
\begin{equation}
    x^2+r(y-\frac{\alpha}{2\beta})^2=1+\frac{\alpha^2}{4\beta}-\cos^2(\theta), 
\end{equation}
where $\beta = 1+ 4\sin^2(2\theta/3)$ and $\alpha=4\sin(2\theta/3)\cos(\theta)$. 
Finally, after defining the new variables $u=\sqrt{r}(y-\alpha/(2\beta))$, $v=x$ and $R=\sqrt{1+\alpha^2/(4\beta)-\cos^2(\theta)}$, we obtain the equation $u^2+v^2=R^2$ which is solved by $u(t)=R\cos(t+t_0)$ and $v(t)=R\sin(t+t_0)$ for some $t_0$. Resubstitution yields 
\begin{equation}
\mathbf{x}_1(t) = \begin{pmatrix} R\sin (t+t_0) \\
R/\sqrt{r}\cos (t+t_0)+2\alpha/\beta \\ -2\sin(2\theta/3)[\mathbf{x}_1]_2(t)+\cos(\theta)
\end{pmatrix}.
\end{equation}
Using the initial condition that $[\mathbf{x}_1]_1(0)=\sin\theta$ we find $t_0=\arcsin[\sin(\theta)/R]$. 
The trajectory corresponding $\mathbf{y}_1(t)$ can be obtained from $\mathbf{x}_1(t)$ by symmetry, see Fig.~\ref{fig:trajectories}, and is given by 
\begin{equation}
\mathbf{y}_1(t) = R_y \mathbf{x}_1(\pi-t),
\end{equation}
where $R_y$ is the reflection along the $y$-axis.

In the Bell scenario, each trajectory is replaced by $d$ states, e.g., the trajectory $\mathbf{x}_i(t)$ corresponds to the states $(\ket{x_{i,1}},\dots,\ket{x_{i,d}})$. To maximize visibility, we choose $\ket{x_{i,k}}$ according to the Bloch vector $\mathbf{x}_i[k\pi/(d+1)]$, such that we have $|\left\langle x_{i,k}|x_{i,k+1}\right \rangle |=| \left\langle w_2|x_{i,1}\right \rangle |=| \left\langle x_{i,d}|w_1\right \rangle|$ for all $1<j<d$. Similarly, we choose the state $\ket{y_{j,k}}$ corresponding to $\mathbf{y}_j[k\pi/(d+1)]$. Finally, we note that the solid angles corresponding to the above states approach the solid angles enclosed by trajectories $\mathbf{x}_i$ and $\mathbf{y}_j$ only for $d\rightarrow \infty$, while for finite $d$, the solid angles correspond to trajectories that connect each adjacent states with geodesics. However, the solid angles for finite $d$ still fulfill $\phi_{\mathrm{g},11}=\phi_{\mathrm{g},12}=\phi_{\mathrm{g},21}=\phi_{\mathrm{g},22}/3$ to a sufficient approximation to lead to nonlocality, see Fig.~\ref{fig:finite_d}.

\end{document}